# Bright light emission and waveguiding in conjugated polymer nanofiberselectrospun from organic-salt added solutions

*Vito Fasano,*[*,†,§] *Alessandro Polini,*[#,,‡] *Giovanni Morello,*[†,#] *Maria Moffa,*[†,§] *Andrea Camposeo,*[*,†,#] *and DarioPisignano*[*,†,§,#]

[†] Center for BiomolecularNanotechnologies @ UNILE, Istituto Italiano di Tecnologia (IIT), Via Barsanti 1, Arnesano (LE), 73010, Italy.

[§] Dipartimento di Matematica e Fisica "Ennio De Giorgi", Università del Salento, via Arnesano, Lecce 73100, Italy.

[#] National NanotechnologyLaboratory of Istituto Nanoscienze-CNR, via Arnesano, Lecce 73100, Italy.

ABSTRACT

Light emitting electrospunnanofibers of poly-[(9,9-dioctylfluorenyl-2,7-diyl)-*co*-(*N,N'*-diphenyl)-*N,N'*-di(p-butyl-oxy-phenyl)-1,4-diaminobenzene)] (PFO−PBAB) are produced by electrospinning under different experimental conditions. In particular, uniform fibers with average diameter of 180 nm are obtained by adding an organic salt to the electrospinning solution. The spectroscopic investigation assesses that the presence of the organic salt does not alter the optical properties of the active material, therefore providing an alternative approach for the fabrication of highly emissive conjugated polymer nanofibers. The produced nanofibers display self-waveguiding of light, and polarized photoluminescence, which is especially promising for embedding active electrospun fibers in sensing and nanophotonic devices.

INTRODUCTION

Nanostructures made of organic semiconductors are attracting a burgeoning interest due to their potential application in micro- and nanoscale photonic and electronic devices such as field effect transistors, light-emitting diodes and photo- or chemical sensors.[1,2] Several studies have shown that organic semiconductor nanofibers and wires show intriguing properties, such as enhanced carrier mobility[3,4] and electrical conductivity,[5-7] and polarized photoluminescence (PL).[8-10] These properties are mostly related to the peculiar arrangement of the polymer backbones, and eventually to the optical transition dipoles within the nanostructures, induced by the reduced transversal size of wires and by the elongating, stretching forces acting on macromolecules during fiber fabrication.[1]

Nanofibers and nanowires made of organic semiconductors have been obtained by different methods, including dip-pen nanolithography,[11,12] self-assembly,[13,14] polymerization in nanoporous templates,[15-17] micro/nanofluidics[18] and electrospinning.[8,19,20] Among these approaches, electrospinning is the most scalable and cost-effective technique allowing ultralong one-dimensional nanomaterials to be synthesized, thanks to its high production yield and relatively cheap equipment,[21-25] even though the industrial upscaling of the process still has open issues.[26] In fact, increasing the number of processable polymers and improving the process reproducibility and accuracy in the production stage are the subject of intense research efforts.[26,27] Different morphologies can be obtained,[28] such as porous,[29] hollow,[30] barbed fibers[31] and necklace-like structures.[32] However, electrospinning of conjugated polymers is still a challenging and non-standardized process due to intrinsic difficulties, related to the polymer chain rigidity, relatively low molecular weight and level of entanglement, and low solubility.[1,19] Some successful approaches exploit the ease-of-processing and favorable plastic behavior of some inert polymers, blended with conjugated polymers.[8,33-37] An elegant method uses two coaxial capillaries to electrospin different liquids in a compound jet. An easily processable polymer solution can be then used to realize the fiber shell, that is removed after electrospinning to obtain pure conjugated polymer fibers.[19,38] Other approaches use an electrospinnable precursor solution and post-processing polymerization.[39-41] For some applications, the availability of nanostructures fully made of conjugated polymers is essential in order to exploit the unique optoelectronic features of $\pi$-conjugated systems. To this aim, effective approaches to electrospin conjugated polymer fibers utilize a mixture of good and poor solvents in order to improve the solution processability.[20,42]

In addition, the processing method may impact the electronic and emissive features of active polymers. The optimization of the resulting light-emitting properties would preferably require the use of good solvents for the conjugated polymers, thus preventing aggregation phenomena that are known to decrease the emission efficiency.[43] Moreover, recent works[10] demonstrate that nanofibers spun by using a single good solvent exhibit a higher molecular alignment and order and, consequently, a higher degree of polarization of the emission. Unfortunately, most of good solvents for conjugated polymers have low boiling point and conductivity,[42,44] strongly disfavoring electrospinning. In fact, efforts to produce conjugated polymer nanofibers by electrospinning from solutions with a single good solvent often lead to leaflike structures,[44] or to fibers with beads[10] or with micrometer diameters.[45] Salts and other additives can be used to increase the solution conductivity without altering significantly the viscosity and surface tension, and this often improves electrospinning performances. This approach allows fibers with regular morphology and ultra-thin diameters (< 10 nm) to be obtained.[46] However, these additives could deteriorate the fiber optical properties, and their effect on conjugated polymer functionality has to be carefully assessed. Though crucial to realize light-emitting nanostructures, this issue is still open for light-emitting polymer nanofibers. While the addition of organic salts such as pyridiniumformate and *p*-toluene sulfonic acid has been investigated for conductive polymers and blends of conjugated polymers withpolysterene and poly(vinyl pyrrolidone) in order to remove the presence of beads and reducethe fiber diameter,[37, 47-49] this method is almost unexplored with nanostructures fully composed by light-emitting conjugated polymers, for which criticalities may be due to the high sensitivity of their emission properties to the composition of thelocal micro-environment, that in turn can induce chain modification by interactions with the solution additives.[50]

In this work we demonstrate the possibility to electrospin smooth, continuous and uniform nanofibers made of the blue light-emitting polymer, poly-[(9,9-dioctylfluorenyl-2,7-diyl)-*co*-(*N,N'*-diphenyl)-*N,N'*-di(p-butyl-oxy-phenyl)-1,4-diaminobenzene)] (PFO-PBAB), by using a single good solvent and a small amount of organic salts. The addition of the organic salts greatly improves the resulting fiber morphology, and, importantly, leaves almost unaltered the PL and spectroscopic properties of the polymer. The process positively affects the waveguiding properties of individual nanofibers as well. These results are therefore very promising for improving the fabrication of functional, conjugated polymer nanofiber building blocks for photonic circuits and optoelectronic applications.

Published in Macromolecules 46:5935-5942, doi: [10.1021/ma400145a](10.1021/ma400145a) (2013).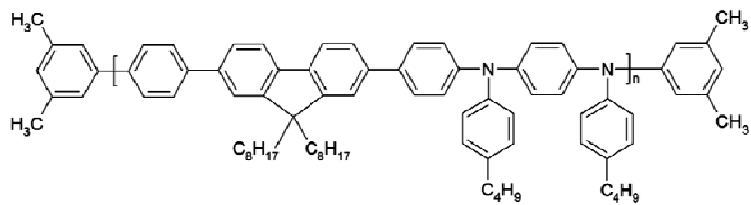

**Scheme 1**. Chemical structure of PFO-PBAB.

EXPERIMENTAL

*Electrospinning.* The chemical structure of PFO-PBAB (American Dye Source, molecular weight = 93 kDa) is shown in Scheme 1. This polymer is used as efficient blue-emitting material in various optoelectronic devices.[51,52] PFO-PBAB is dissolved in chloroform with a concentration of 120 mg/mL and either tetrabutylammonium iodide (TBAI, Sigma Aldrich) ortetrabutylammonium bromide (TBAB, Lancaster) are added under stirring and ultrasonic bath [PFO-PBAB:TBAI(TBAB) 10:1, w:w].

The polymer solution is loaded in a syringe with a 27 gauge stainless steel needle, and a 5 kV voltage is applied to the needle by a power supply (Glassman High Voltage). Quartz coverslips or Al foils are placed at a distance of 20 cm from the needle on a 10×10 cm$^2$ collector, negatively biased (-6 kV). Electrospinning is performed with an injection flow rate of 5 μL/min, and a relative humidity and temperature of about 60% and 22°C, respectively. Alternatively, PFO-PBAB fibers are produced by dissolving the polymer in a mixture of tetrahydrofuran (THF) and dimethyl sulfoxide (DMSO, 9:1 v/v) with a polymer concentration of 120 mg/mL. Uniform fibers are obtained by negatively biasing the collector (- 6 kV), by applying a positive bias of 5 kV to the needle, with a flow rate of 8 μL/min and the collector at a distance of 10 cm from the needle. Using a single solvent (CHCl$_3$ and THF) for dissolving the polymer has the main effect of drastically increasing the density of beads (inset of Fig. 1a) for any combination of the other

Published in Macromolecules 46:5935-5942, doi: 10.1021/ma400145a (2013).process parameters. For polarized infrared spectroscopy, freestanding arrays of uniaxially aligned nanofibersare fabricated by a collector (a disk with diameter of 8 cm and thickness of 1 cm) rotating at 4000 rpm, positioned at a distance of 10 cm from the needle.

Reference thin films are realized by spin-coating at 6000 rpm. Films and fibers with comparable thickness are selected for optical investigation, in order to minimizeartifacts due to self-absorption. Before experiments, samples are stored in vacuum at room temperature for at least one night to remove solvent residues.

*Morphological and spectroscopic measurements.* The morphology of fibers is investigated by scanning electron microscopy (SEM, FEI Nova NanoSEM 450) operating at 5–10 kV. Ultraviolet-visible (UV-Vis) absorption spectra of thin-films are collected by using a spectrophotometer Varian Cary 300 Scan). Polarized optical maps of electrospun PFO-PBAB nanofibers are obtained by a microscope (Olympus, BX52) equipped with a Hg fluorescence lamp, a 50× objective (Olympus, UMPlan FL, NA = 0.75), a rotating polarized filter and a remotely-controlled CCD camera. PL spectra are measured by using a spectrometer (Ocean Optics USB 4000), exciting samples by a cwdiode laser ($\lambda$=405 nm). The absolute quantum efficiency ($\phi$) of films and fibers is obtained by exciting samples in an integrating sphere (Labsphere) by the diode laser and analyzing PL by a fiber-coupled spectrometer. All the spectra are corrected by the spectral response of the experimental setup (integrating sphere, optical fiber and spectrometer).The FTIR spectra are acquired with a spectrometer (Vertex 70, Bruker) and a IR grid polarizer (Specac Limited, U.K.), consisting of 0.12 m wide strips of aluminum, mounted on a rotation stage. The 8 mm wide beam, incident orthogonally to the plane of the sample, is polarized parallel, orthogonal or at variable angle with respect to the main alignment axis of fibers.

Confocal fluorescence maps are obtained by a laser scanning microscope (Nikon A1R-MP equipped with spectral scan head). The confocal system consists of an inverted microscope (Eclipse Ti, Nikon), an oil immersion 60× objective (NA=1.40, Nikon) and an excitation laser source ($\lambda$ = 408 nm). The emission is collected through the microscope objective and the intensity is measured by a spectral detection unit equipped with a multi-anode photomultiplier.

The waveguiding properties of electrospunnanofibers are analyzed by using a micro-photoluminescence ($\mu$-PL) setup, based on an inverted microscope (IX71, Olympus) equipped with a 60× oil immersion objective (NA=1.42, Olympus) and a CCD camera. The PL is excited by the diode laser coupled to the microscope through a dichroic mirror and focused on the sample by the objective. Part of the light emitted by the conjugated polymer, excited by the tightly focused laser spot, is coupled into the nanofiber and waveguided. The fiber optical losses coefficient is measured acquiring an image of the intensity of emission diffused by the fiber surface, and analyzing the spatial decay of emission as a function of the distance from the exciting laser spot.[20] Finally, time-resolved PL measurements are performed in single-photon counting mode by exciting the samples at a low excitation level at $\lambda$ = 338 nm with a repetition rate of 1 kHz.

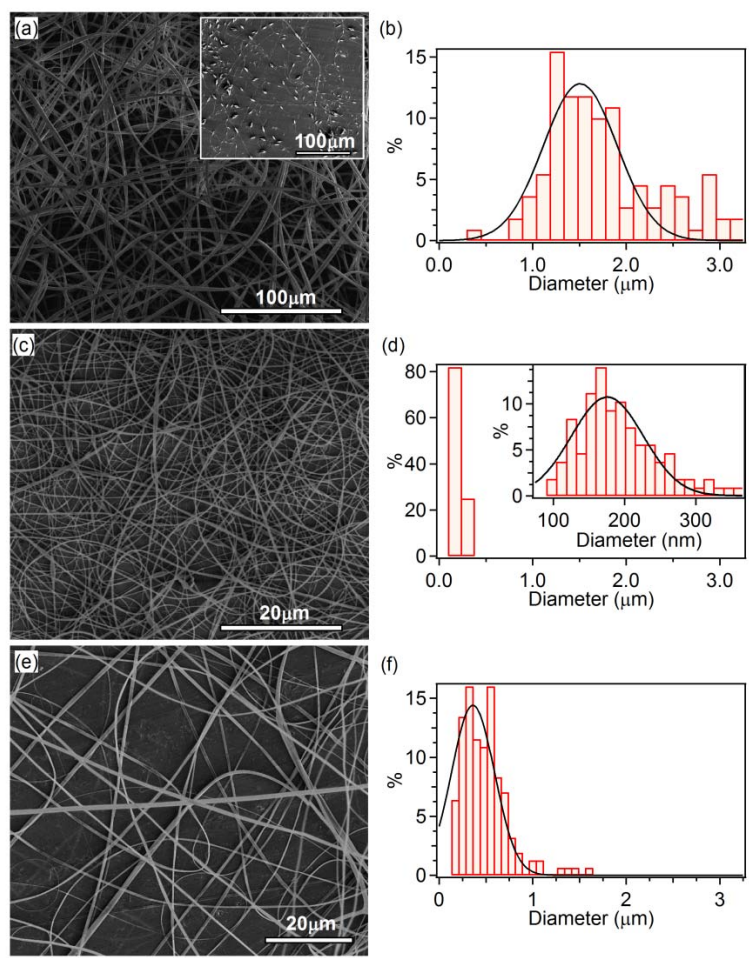

**Figure 1.** (a)–(b) SEM micrograph and fiber diameter distribution of PFO-PBAB electrospun fibers obtained by using a THF:DMSO mixture (scale bar = 100 μm). Inset: SEM image of PFO-PBAB fibers fabricated by using a single solvent (CHCl$_3$, scale bar = 100 μm). (c)-(f) SEM micrographs [(c) and (e)] and fiber diameter distribution [(d) and (f)] of electrospun PFO-PBAB fibers obtained dissolving the conjugated polymer in CHCl$_3$ with the addition of TBAI and TBAB, respectively (scale bar = 20 μm). Continuous lines in (b), (d) and (f) are Gaussian fits to the data.

RESULTS AND DISCUSSION

In Figure 1we display the SEM micrographs and analysis of PFO-PBAB electrospun fibers obtained from different solutions. The inset of Figure 1a shows fibers obtained by electrospinning from a single good solvent (chloroform), evidencing the presence of abundant and large beads along the fibers. Uniform and continuous fibers can be instead obtained by using a mixture of good and poor solvents,[20,42] namely tetrahydrofuran (THF) and dimethylsulfoxide (DMSO), respectively (9:1 v/v, Figure 1a-b).

However, the average diameter of these fibers is still around 1.5 μm, and trying to fit the diameter distribution by a Gaussian curve leads to a standard deviation, $\sigma$, as high as 600 nm (Figure 1b). In Figure 1c-d, we display a SEM micrograph and the analysis of fibers electrospun by adding the TBAI organic salt to the PFO-PBAB/chloroform solution. In this way the bead-structure of Figure 1a is completely absent and the resulting continuous, smooth and uniform PFO-PBAB nanofibers have an average diameter of 180 nm and $\sigma$ of 70 nm (inset in Figure 1d). These values are significantly smaller than in other reported pristine conjugated polymer nanofibers, having typical average diameter > 200 nm and larger dispersions in size (> 100 nm).[20, 42] In addition, Figure 1e-f showsa SEM micrograph and the corresponding diameter distribution of electrospunfibers obtained from a PFO-PBAB/TBAB chloroform solution, at optimized electrospinning conditions. The average diameter of thefibers is about 360 nm ($\sigma$= 320 nm), larger than the values obtained by using the TBAI salt.

Organic salts are often used for improving electrospinnability and nanofibers uniformity, especially with optically inert polymers.[47,53-55] Indeed, this results in a higher charge density and ultimately in higher elongation forces experienced by the jet. The diameter of the obtained electrospun fibers also becomes essentially smaller,[53,54] which is also consistent with models predicting a decrease of the terminal radius, $h_t$, of electrospun jets upon increasing the solution conductivity.[56] In order to investigate the impact of the addition of the organic saltson the optical properties of PFO-PBAB fibers, we firstly characterize the absorption and PL of spin-coated thin films (Figure 2a). The absorption spectrum features a peak at 375 nm, with similar values of the full width at half maximum (FWHM) and of the maximum absorption coefficient for the pristine PFO-PBAB samplesand for salt-added samples (Table 1).

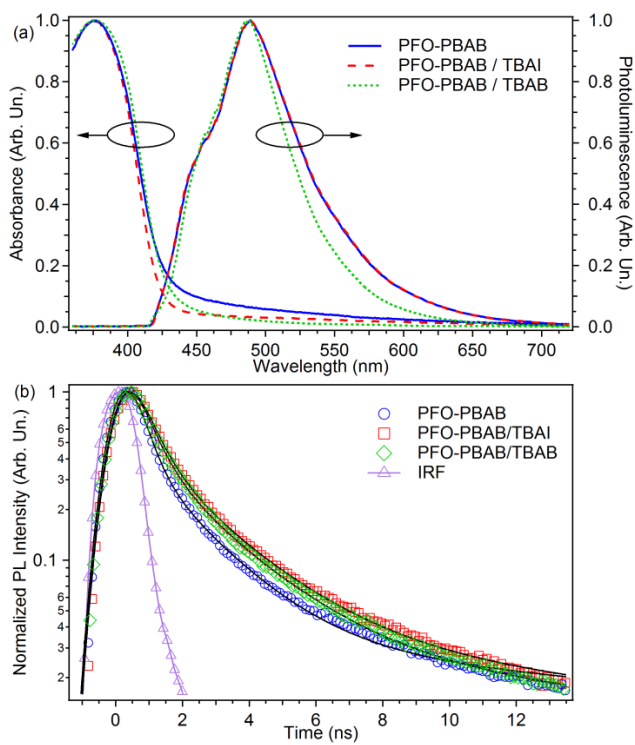

**Figure 2.** (a) Normalized absorption and PL spectra of spin-coated films of pristine PFO-PBAB film (blue continuous lines) and of PFO-PBAB with TBAI (dashed line) and TBAB (dotted line), respectively. (b) Time-profiles of PL decay of a pristine PFO-PBAB film (circles) and of PFO-PBAB with TBAI (squares) and TBAB (diamonds). The instrument response function is also shown (triangles). The black continuous lines are fits to the data by a sum of three exponential functions convoluted with the IRF.

PL is almost unchanged as well, with only a small decrease of the FWHM (about 10 nm) in the samples with TBAB being observed. Time-resolved PL measurements highlight a clear non exponential fluorescence decay, evidencing the presence of different emitting species and the existence of multiple electronic states (Figure 2b). A detailed analysis of such emissive species is beyond the scope of the present paper. However, the decay data can be fitted by the sum of three exponential functions, convoluted with a Gaussian function to account for the instrument response function (IRF).[57,58]

**Table 1.** Spectroscopic properties of PFO-PBAB spin-coated films without and with TBAI or TBAB.

|  | PFO-PBAB | PFO-PBAB/TBAI | PFO-PBAB/TBAB |
|---|---|---|---|
| Abs $_{max}$ (nm) | 375±1 | 375±1 | 375±1 |
| Abs FWHM (nm) | 65±1 | 66±1 | 68±1 |
| $\varepsilon_{max}$ (cm$^{-1}$) | (1.5±0.2)×10$^5$ | (1.3±0.1)×10$^5$ | (1.3±0.1)×10$^5$ |
| PL $\lambda_{max}$ (nm) | 489±1 | 488±1 | 488±1 |
| PL FWHM (nm) | 88±1 | 88±1 | 77±1 |
| $\phi$ | 0.13±0.01 | 0.13±0.01 | 0.14±0.01 |
| <$\tau_{PL}$> (ns) | 1.3±0.2 | 1.4±0.2 | 1.4±0.1 |

In order to compare the emission performances of the investigated samples we consider an amplitude–weighted lifetime, given by $<\tau> = \sum_{i=1}^{3} A_i \tau_i$, where $A_i$ is the normalized amplitude of the *i*-th exponential component. The results (Table 1) evidence comparable amplitude-weighted lifetimes. Moreover, the measured absolute quantum efficiencies of the reference thin films are also almost identical (13-14%, Table 1). Overall, the presence of the organic salts does not alter significantly the fluorescence properties of PFO-PBAB films.

In electrospun fibers as well, confocal fluorescence imaging evidences a bright and uniform PL intensity along the longitudinal axis of the nanostructures (Figure 3a-b). In Figure 3c-d, we compare the PL spectrum of a mat of PFO-PBAB fibers with that of the corresponding film. The PL spectrum of the fibers made by adding the TBAI salt ($\lambda_{max}$ = 490 nm, FWHM = 78 nm, Fig. 3c) shows a slight decrease of the linewidth compared to the reference film ($\lambda_{max}$ = 488 nm, FWHM = 88 nm), mainly due to the difference of the intensity of the high energy shoulder of the PL spectrum, likely due to a vibronic replica. This difference is mainly attributed to the residual self-absorption, because of the thickness of the analyzed fibers mats which is less uniform then in the film. Fibers produced by adding the TBAB salt do not show significant differences compared to the corresponding film (Fig. 3d). Similar results are found for fibers made without adding the salts, both with single solvent and by the investigated solvent mixture (see Supporting Information). Overall, the largely unperturbed emission properties of PFO-PBAB under the different processing conditions make this material particularly suitable for nanophotonic applications.

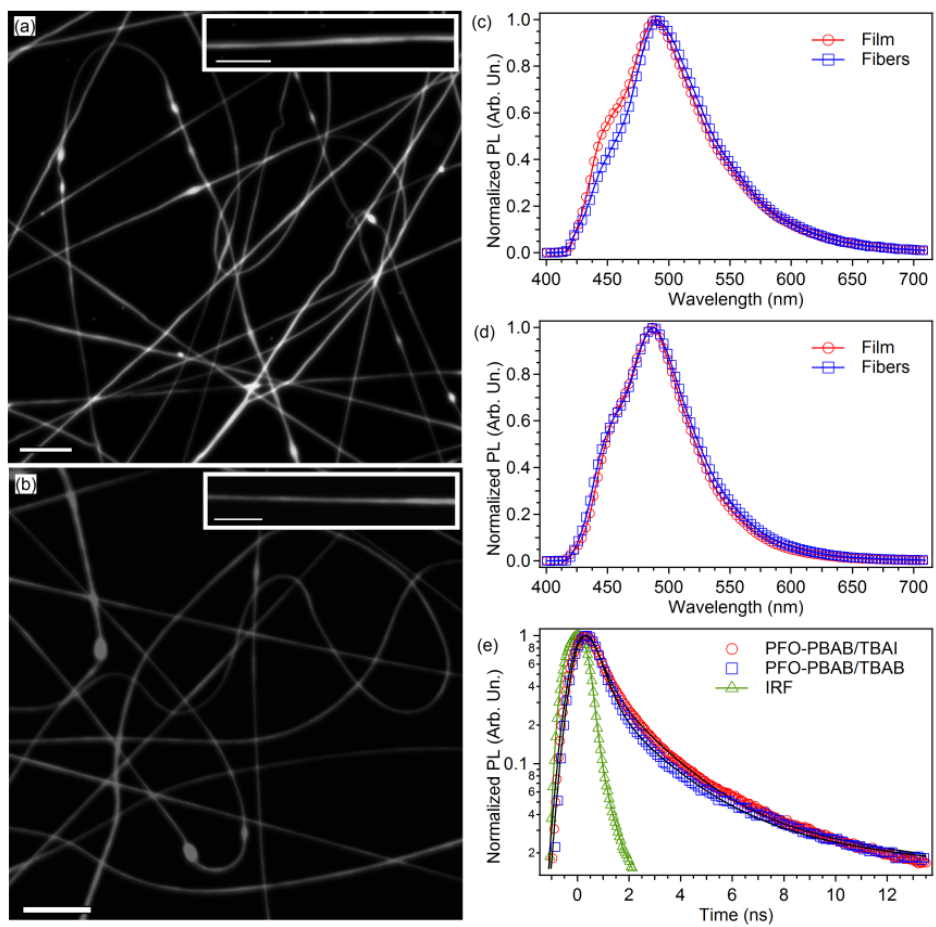

**Figure 3.** (a, b) Fluorescence confocal micrographs of a mat of PFO-PBAB/TBAI (a) and PFO-PBAB/TBAB (b) fibers (scale bar = 10 μm). Examples of individual light-emitting nanofibersareshown in the corresponding insets(scale bars = 5 μm). (c) PL spectra of PFO-PBAB nanofibers (circles) and films (squares) with TBAI. (c) PL spectra of PFO-PBAB nanofibers (circles) and films (squares) with TBAB. (d) PL temporal decay for PFO-PBAB/TBAI(circles) and PFO-PBAB/TBAB nanofibers (squares).The black continuous linesare the best fit to the data by a sum of three exponential functions convoluted with the IRF (the latter is also shown with triangles).

The time decay profiles of the PFO-PBAB nanofibers PL areshown in Figure 3e. Compared to the reference films, the overall decays of the nanofiber emission are faster, and the amplitude–weighted lifetime obtained by fitting is about 1 ns. As for films, data are well fitted by the sum of three exponential functions convoluted with the IRF function. A minor shortage (~10%) is found for the three contributing components compared to film values, an effect attributable to the more ordered packing of the PFO-PBAB macromoleculesinto the fibers (see below).

Conjugated polymer nanofibers can also be exploited as active waveguides.[20] To assess the propagation losses of light guided in PFO-PBAB fibers, the intensity of the PL escaping from the fiber surface and tip is imaged by $\mu$-PL (Figure 4a) and measured as a function of the distance from the excitation spot, $d$. Figure 4b shows typical PL images collected at different values of $d$, evidencing effective waveguiding of the light excited by the focused laser beam. These images are acquired on a freestanding nanofiber made by adding the TBAI salt, and having subwavelength size. Waveguidingis clearly appreciable for distances up to 0.2mm and also in bent fibers (inset of Fig. 4c).These data allow us to estimate the loss coefficient, $\alpha$, which is of the order of 100 cm$^{-1}$, i.e. much lower than values typically measured in active conjugated polymer nanofibers.[15,20,59] Higher values of the loss coefficient, ranging from 700 cm$^{-1}$ to 2000 cm$^{-1}$, are measured for fibers deposited on quartz substrates (Fig. 4c), which is attributable to a partial coupling of guided lightinto the substrate, mainly by evanescent fields. In fact the fraction of power of the fundamental mode of a cylindrical waveguide, $\eta$, depends on the diameters of the guide as:[59]

$$\eta = 1 - \left[\frac{(2.4e^{-1/V})^2}{V^3}\right], \text{ where } V = \frac{\pi d_{fiber}}{\lambda}\sqrt{n_{fiber}^2 - n_0^2} \quad (1)$$

In the above expression, $d_{fiber}$ is the nanofiber diameter, and $n_{fiber}$ and $n_0$ are the refractive index of the fiber (about 1.8) and of the surrounding medium, respectively. The dependence of $\eta$ on $d_{fiber}$ is shown in the inset of Fig. 4c for a waveguide in air. Due to their reduced size, the fibers produced with the TBAI salt ($\eta$=85%) are most sensitive to variations of their environment, producing a change of the refractive index and, consequently, a perturbation of the field into the waveguide.

The measured loss coefficients are comparable to those reported for other conjugated polymer fibers,[15,20] in which optical losses are typically associated to self-absorption and scattering from surface and bulk defect or inhomogeneities. Interestingly, estimating the contribution of self-absorption by the absorption spectra measured for thin films (Figure 2a), we find a significantly higher expected value of $\alpha$ ($6\times10^3$ cm$^{-1}$). This discrepancy can be related to a preferential supramolecular organization and orientation of the polymer backbones induced by electrospinning.[8,10,60] Indeed, this effect can lower the self-absorption of the guided light, whose wavevector would be parallel to the transition dipole moment of the molecules oriented along the fiber axis, thus ultimately reducing photon re-absorption.

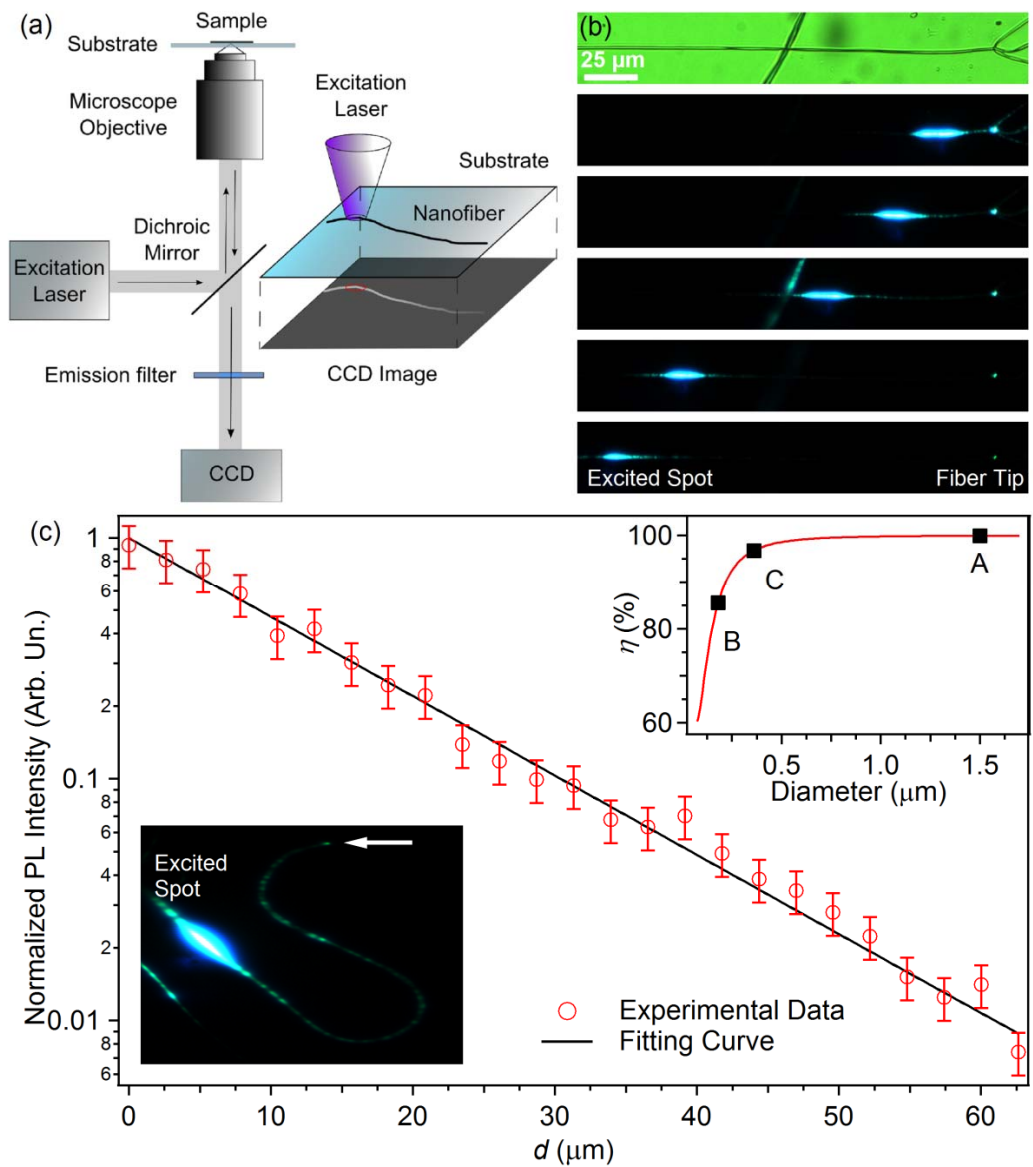

**Figure 4.** (a) Scheme of the experimental setup used for the characterization of single-fiber waveguiding. (b) Images of a fiber excited by a focused laser beam, positioned at a variable distance from the fiber tip. The top panel is a brightfield image of the investigated nanofiber. (c) Spatial decay of the light intensity (red circles) guided along a single electrospun fiber, deposited on a quartz substrate, as function of distance, $d$, from the photoexcitation spot. The continuous line is the best fit of the experimental data by an exponential function, $I=I_0\exp(-\alpha d)$. Bottom-left

inset: micrograph showing light guided in a bentactive polymer fiber. The horizontal arrow highlights the fiber tip, whereas the bright spot corresponds to the emission directly excited by the focused laser beam. Right-top inset: plot of the fraction of guided power in the fundamental mode as a function of the fiber diameter, calculated by using Eq. (1). Points labeled as A, B, C correspond to the average diameter of fibers fabricated by using a THF:DMSO mixture of solvents (A), and by the addition of TBAI (B) and TBAB (C), respectively. The morphology of these fibers shown in Fig. 1.

To probe the orientation of the molecules within the electrospunnanostructures, polarized FTIR absorption spectroscopy is performed on freestanding uniaxially aligned arrays offibers. Spectra collected with the incident light polarization parallel and perpendicular to the fiber axis are shown in Figure 5a, evidencing a preferential absorption of light polarized along the fiber length. In particular, by considering the peak at 1603 cm$^{-1}$ (inset of Fig. 5a), attributed to the ring stretching mode of the fluorene unit, that is associated to vibrations prevalently directed along the molecular chain axis,[10,61] a dichroic ratio (ratio between the absorbance of light polarized parallel to the fiber axis and light polarized perpendicularly to the fiber axis) of about 2 is measured. This is shown in Figure 5b, where the intensity of the 1603 cm$^{-1}$ peak is displayed as a function of the angle between the direction of polarization of the incident infrared light and the axis of alignment of thefibers. This result clearly indicates the preferential alignment of the polymer chains along the fiber axis.

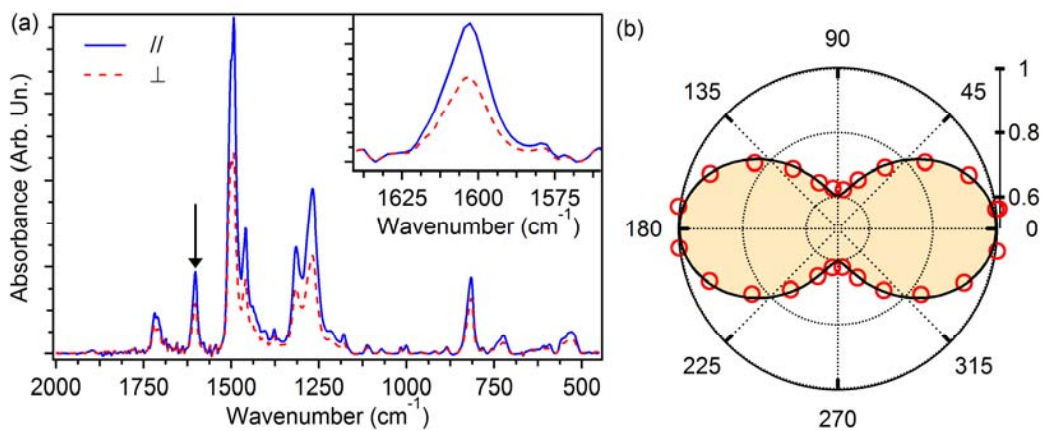

**Figure 5.** (a) Polarized FTIR absorption spectra of freestanding mats of aligned electrospun fibers realized by the addition of TBAI. The spectra are acquired with incident infrared light polarized parallel (continuous line) and perpendicular (dashed line) to the fiber axis. The inset shows the peak at 1603 cm$^{-1}$ utilized for the analysis and highlighted by an arrow in the main panel. (b) Absorbance vs. the angle formed by the fiber axis and the polarization of the incident light. Data, obtained for the mode at 1603 cm$^{-1}$, are normalized to the value of maximum absorbance, measured for polarization of the incident light parallel to the fibers.

The orientation of optical transition dipoles in individual PFO-PBAB fibers can be probed also by polarized emission microscopy. Polarized fluorescence micrographs (Figure 6a-c) evidence a variation of the intensity as function of the angle between the polarizer filter axis and the nanofiber longitudinal axis. The resulting PL polarization ratio ($\chi=I_{\parallel}/I_{\perp}$) is about 2, which confirms a preferred alignment of polymer backbones along the fiber length. The here found polarization ratio is comparable to that reported for other light-emitting electrospun systems.[8,10,60] Similar measurements (data not shown) performed on spin-coated films evidence unpolarized emission. The intrinsic alignment of polymer macromolecules along the fiber axis, hence of emissive transition dipoles, may cause the relatively low values of propagation losses measured in PFO-PBAB nanofibers compared to linear attenuation coefficient estimated from films data. Reduced self-absorption makes these blue-emitting fibers promising for use in miniaturized photonic sensors and devices.

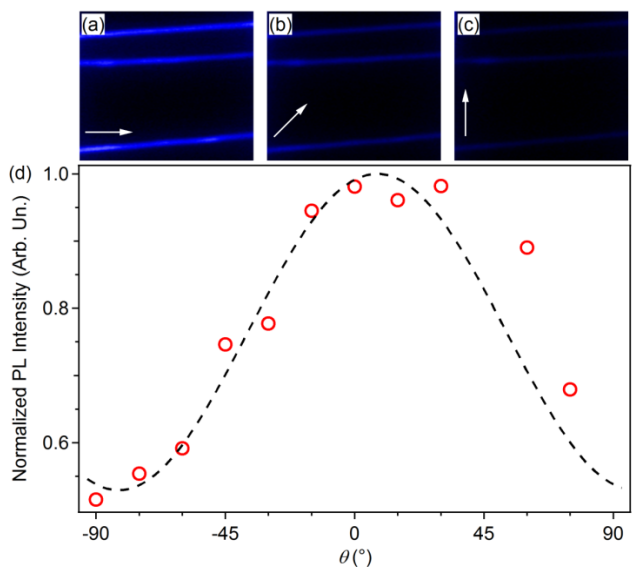

**Figure 6.** (a)-(c). Micrographs of the emission intensity of individual PFO-PBAB nanofibers, with different position of the analyzer. The angles, $\theta$, formed by the fiber and the analyzer axes in (a), (b), and (c) are 0°, 45° and 90°, respectively. The analyzer axis direction is highlighted with white arrows. Here excitation is carried out by the unpolarized light of a Hg lamp, coupled into a microscopy objective. (d) Emission intensity as function of the angle between the analyzer and the fiber axes. The dashed line is a fit to the data by the Malus law $I = I_0 + I_1 cos^2\theta$, where $I_0$ indicates the intensity of the unpolarized background. Obtained parameters are $I_0$=(0.55±0.05) and $I_1$=(0.5±0.1).

CONCLUSIONS

Continuous and uniform, bright blue light-emitting fibers can be realized by electrospinning a conjugated polymer (PFO-PBAB) using a single good solvent with the addition of organic salts (TBAI and TBAB). The addition of organic salts to the electrospinning solution is effective for promoting the formation of uniform fibers with no bead-like structures. Individual nanofibers realized by this approach have active waveguiding characteristics and polarized PL, whose features are almost unaltered with respect to samples obtained without the organic salts. In perspective, these fibers can be used as optically active elements for sensing and photonics and in light-emitting optoelectronic devices.

ASSOCIATED CONTENT

**Supporting Information**. A supporting document with additional technical details on electrospinning process and nanofibers emission spectra is included as a separate PDF file. This material is available free of charge via the Internet at http://pubs.acs.org.

AUTHOR INFORMATION

**Corresponding Author**

* Vito Fasano: vito.fasano@iit.it

* Andrea Camposeo: andrea.camposeo@nano.cnr.it

* Dario Pisignano: dario.pisignano@unisalento.it

**Present Addresses**

‡ Center for Biomedical Engineering, Department of Medicine, Brigham and Women's Hospital, Harvard Medical School, Cambridge, USA

ACKNOWLEDGMENTS

The research leading to these results has received funding from the European Research Council under the European Union's Seventh Framework Programme (FP/2007-2013)/ERC Grant Agreement n. 306357 (ERC Starting Grant "NANO-JETS"). The Apulia Regional Network of Public Research Laboratories N. 09 (WAFITECH) is also acknowledged for SEM measurements.

Published in Macromolecules 46:5935-5942, doi: [10.1021/ma400145a](10.1021/ma400145a) (2013).(8)   Campoy-Quiles, M.; Ishii, Y.; Sakai, H.; Murata, H. *Appl. Phys. Lett.* **2008**, *92*, 213305.

(9   Kuo, C. C.; Wang, C. T.; Chen, W. C. *Macromol. Mater. Eng.* **2008**, *293*, 999.

(10)   Pagliara, S.; Vitiello, M. S.; Camposeo, A.; Polini, A.; Cingolani, R.; Scamarcio, G.; Pisignano, D. *J. Phys. Chem. C* **2011**, *115*, 20399.

(11)   Noy, A.; Miller, A. E.; Klare, J. E.; Weeks, B. L.; Woods, B. W.; De Yoreo, J. J. *Nano Lett.* **2002**, *2*, 109.

(12)   Maynor, B. W.; Filocamo, S. F.; Grinstaff, M. W.; Liu, J. *J. Am. Chem. Soc.* **2002**, *124*, 522.

(13)   Liu, J.; Sheina, E.; Kowalewski, T.; McCullough, R. D. *Angew. Chem. Int. Edit*. **2002**, *41*, 329.

(14)   Samitsu, S.; Shimomura, T.; Heike, S.; Hashizume, T.; Ito, K. *Macromolecules* **2008**, *41*, 8000.

(15)   Carroll, D.; Lieberwirth, I.; Redmond, G. *Small* **2007**, *3*, 1178.

(16)   Steinhart, M.; Wendorff, J. H.; Greiner, A.; Wehrspohn, R. B.; Nielsch, K.; Schilling, J.; Choi, J.; Gösele, U. *Science* **2002**, *296*, 1997.

(17)   Lee, K. J.; Oh, J. H.; Kim, Y.; Jang, J. *Adv. Mater.* **2006**, *18*, 2216.

(18)   De Marco, C.; Mele, E.; Camposeo, A.; Stabile, R.; Cingolani, R.; Pisignano, D. *Adv. Mater.* **2008**, *20*, 4158.

(19)   Li, D.; Babel, A.; Jenekhe, S. A.; Xia, Y. *Adv. Mater.* **2004**, *16*, 2062.
23

# Supporting Information

# Bright Light Emission and Waveguiding in Conjugated Polymer NanofibersElectrospun from Organic Salt Added Solutions

*Vito Fasano,*[*,†,§] *Alessandro Polini,*[#,,‡] *Giovanni Morello,*[†,#] *Maria Moffa,*[†,§] *Andrea Camposeo,*[*,†,#] *and Dario Pisignano*[*,†,§,#]

[†] Center for BiomolecularNanotechnologies @ UNILE, Istituto Italiano di Tecnologia (IIT), Via Barsanti 1, Arnesano (LE), 73010, Italy.

[§] Dipartimento di Matematica e Fisica "Ennio De Giorgi", Università del Salento, via Arnesano, Lecce 73100, Italy.

[#] National NanotechnologyLaboratory of Istituto Nanoscienze-CNR, via Arnesano, Lecce 73100, Italy.

AUTHOR INFORMATION

**Corresponding Author**

* Vito Fasano: vito.fasano@iit.it

* Andrea Camposeo: andrea.camposeo@nano.cnr.it

* Dario Pisignano: dario.pisignano@unisalento.it

**Present Addresses**

‡ Center for Biomedical Engineering, Department of Medicine, Brigham and Women's Hospital, Harvard Medical School, Cambridge, USA.

1. **Electrospinning**

The optimal electrospinning process variables have been selected by systematically varying the polymer concentration in the range 80-120 mg/mL, the applied voltage in the range 2-30 kV, the flow rate between 2 and 20 μL/min and the needle-collector distance in the interval 10-20 cm. The samples obtained at the various set of process parameters are imaged by optical and scanning electron microscopy (SEM), in order to find the set providing the optimal fiber morphology, namely absence of beads and reduced width of the distributions of diameters. In Figure S1 we show exemplary SEM images of fibers of PFO-PBAB obtained by changing the applied voltage in the range 5-25 kV. For fibers obtained by adding TBAI, we typically find a slight increase of the fiber diameters and of the density of beads upon increasing the electric field, whereas for fibers spun from the solvent mixture (THF:DMSO) similar morphological changes are observed by decreasing the electric field. The decrease of the flow rate typically leads to an increase of the density and size of beads. Electro-spray is observed upon decreasing the polymer concentration by 30%.

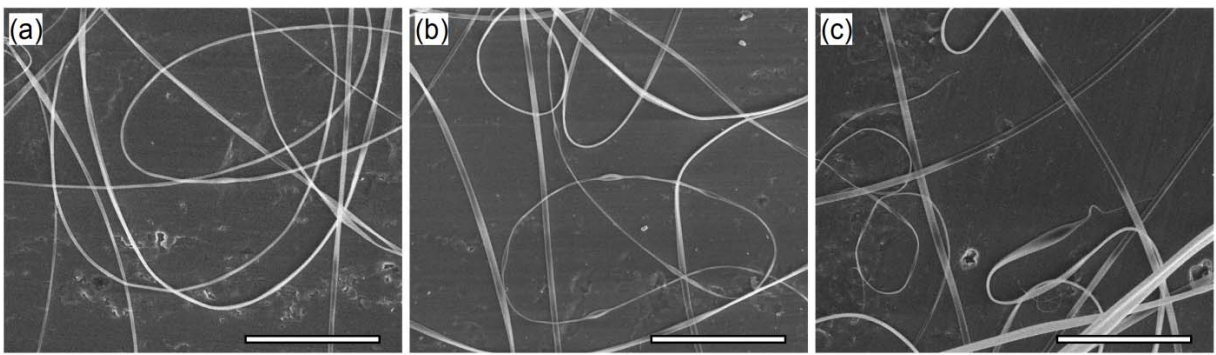

**Figure S1.** (a)–(c) SEM micrographs of electrospun PFO-PBAB fibers fabricated by using a single solvent (120 mg/mL PFO-PBAB/CHCl$_3$) and TBAI organic salt (PFO-PBAB/TBAI 10:1 w:w) at different voltages applied to the needle. (a) 5kV; (b) 20kV; (c) 25kV. Scale bars = 10 μm. These experiments are performed by biasing the collector at -6 kV, by using a flow rate of 5 mL/min and with a needle-collector distance of 20 cm.

2. **Nanofiber emission spectra**

In Figure S2 we compare the spectra of the PFO-PBAB fibers obtained under the optimized electrospinning conditions as reported in the experimental section. All the spectra are peaked at about 490 nm and have a FWHM of about 80 nm. The small differences of the emission lineshape are attributed to the different sample morphology, which can affect the emission spectral shape by scattering and self-absorption.[S1] Fibers obtained by electrospinning PFO-PBAB dissolved in the THF:DMSO solvent mixture display a narrower emission spectrum (FWHM =

73 nm). We attribute this difference to the fiber size that, being of the order of few microns, favours self-absorption effects and a decrease of the intensity of the high energy components of the spectrum, as shown in Fig. S2.

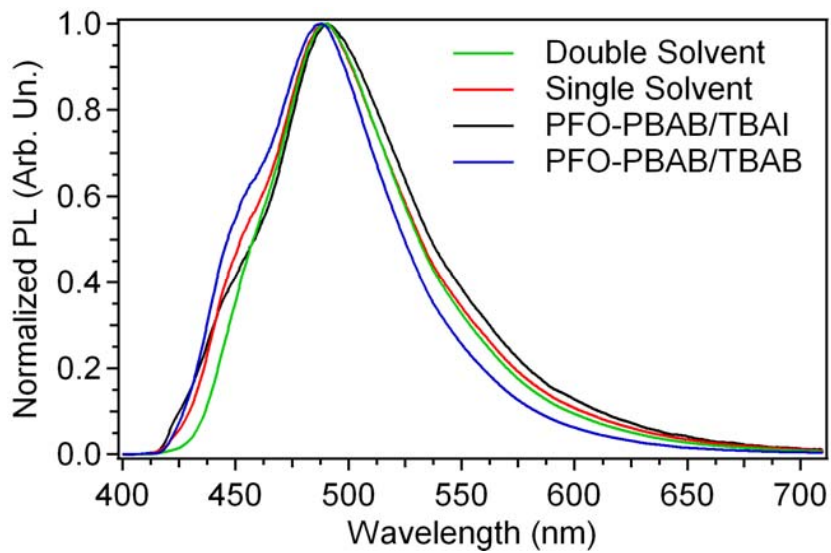

**Figure S2.** Comparison of the emission spectra of the PFO-PBAB fibers electrospun under different conditions: single solvent ($CHCl_3$, red line), THF:DMSO solvent mixture (green line), $CHCl_3$ with TBAI (black line) and $CHCl_3$ with TBAB (blue line). The morphology of the fibers used for the PL measurements is shown in Fig. 1 in the paper.